# PURE: A Framework for Analyzing Proximity-based Contact Tracing Protocols

FABRIZIO CICALA*, WEICHENG WANG, TIANHAO WANG, NINGHUI LI, ELISA BERTINO, FAMING LIANG, and YANG YANG

Many proximity-based tracing (PCT) protocols have been proposed and deployed to combat the spreading of COVID-19. In this paper, we take a systematic approach to analyze PCT protocols. We identify a list of desired properties of a contact tracing design from the four aspects of Privacy, Utility, Resiliency, and Efficiency (PURE). We also identify two main design choices for PCT protocols: *what information patients report* to the server, and *which party performs the matching*. These two choices determine most of the PURE properties and enable us to conduct a comprehensive analysis and comparison of the existing protocols.

CCS Concepts: • **Security and privacy** → **Privacy-preserving protocols**; • **Applied computing** → **Health care information systems**.

Additional Key Words and Phrases: Contact Tracing, COVID-19

## 1 INTRODUCTION

The COVID-19 pandemic is shaping up to be among the biggest disasters for humanity since World War 2 [46]. As of November, 2020, over one million people have died due to it, and the eventual fatality number will likely be in the millions. The financial damage is going to be trillions of dollars [14, 63]. One proven tool we can use to fight against highly infectious diseases like COVID-19 is contact tracing, and case investigation [2]. To trace disease transmission, public health personnel (e.g., *contact tracers*) work with people who have been tested positive (which we call patients) to help them remember everyone with whom they have had close contact during the time they may have been infectious. Exposed persons (contacts) are then notified of their potential exposure and possibly quarantined if tested positive. Contacts are provided with education, information, and support to help them understand their risk, what they should do to separate themselves from others who are not exposed, and how to monitor themselves for illness [2].

Traditional contact tracing requires extensive manual efforts and scales poorly, as contact tracers need to interview patients to identify contacts. Furthermore, some contacts, such as those in public transportation and public areas like airplanes and bars, are inherently difficult to identify. These factors led some experts to conclude early on that containment of COVID-19 has failed, and the society should move on to the mitigation phase [60]. However, the tremendous human and economic damage caused by COVID-19 led all societies to take dramatic measures to help curtail or at least slow down the transmission of COVID-19.

Technologies can potentially help automate the contact tracing process, reducing the cost and improving accuracy at the same time. A wide range of technologies have been applied. For example,

---

*The first two authors contributed equally to this research.

Authors' address: Fabrizio Cicala, fcicala@purdue.edu; Weicheng Wang, wang3623@purdue.edu; Tianhao Wang, tianhaowang@purdue.edu; Ninghui Li, ninghui@cs.purdue.edu; Elisa Bertino, bertino@purdue.edu; Faming Liang, fmliang@purdue.edu; Yang Yang, yangyang@ufl.edu.

some countries use technologies to ensure that quarantine rules have been followed. Patients in self-quarantine are required to wear specific devices (e.g., smart bracelets in Hong Kong [70]), which monitor the movement of the patients. Once the movement exceeds a given limit, the health personnel will be alerted to intervene. South Korea uses CCTV video footage and financial transaction records for contact discovery [73]. Officials in South Korea identify infected people's movements based on card transaction data submitted by 22 credit card operators [44].

As smartphone usage is ubiquitous in modern society, the use of smartphone apps to help with contact tracing and case management is a general, easily deployable solution. Smartphone apps that help contact tracing have already been deployed in several countries, including Iceland, Singapore, Australia, Saudi Arabia, and many others [31, 39, 40, 80]. Mobile apps-based solutions can be categorized into two broad approaches: location-based and proximity-based. [81]

In location-based contact tracing, a central server obtains the locations of each phone over time and uses this information to identify contacts between patients and other users. Such location information can be obtained from the cellular service provider or reported by apps that use GPS on smartphones. For such systems to work, the server generally obtains movement traces of all users, creating a new surveillance capability that does not exist in traditional contact tracing.

In **Proximity-based Contact Tracing (PCT)** systems, the goal is to identify events where two persons are physically close to each other, by detecting that their smartphones are physically close for some period of time. PCT relies on the availability of short-distance communication capabilities such as Bluetooth Low Energy (BLE) [22] and ultrasonic technologies on smartphones. Using such communication capabilities, phones can exchange **beacons** as well as estimate the distance to each other. In PCT, physical proximity can be detected without revealing the absolute physical locations.

The security community quickly recognized that expertise in cryptographic protocols and security analysis can be used to design PCT protocols that support effective contact tracing and privacy at the same time. Many proposals have thus been introduced, e.g., Apple/Google protocols [12], ROBERT [26], DP3T [35], Bluetrace [17], Tracesecure [18], Pronto C2 [16], Epic [10], Epione [76], DESIRE [27]. There are also analysis and debates over which protocols provide a better solution [6, 29, 50, 52, 54, 56, 66, 68, 74, 79].

In this paper, we introduce a framework to analyze PCT protocols. We identify two main dimensions, along with which different designs for PCT protocols can be made. The first is *what information patients report* to the server, to enable the discovery of exposure events. Patients can report what they **sent**, what they **received**, or some secrets **agreed** between them and other users in potential exposure events. The second dimension is *which party performs the matching* between information reported by patients and users' information. Matching could be performed by the **server**, by each individual **user**, or by the server and each user jointly through some **interactive** protocol. These two dimensions thus yield nine possible combinations. We found that all PCT protocols we have examined fall within one of the nine combinations, and this combination determines most (but not all) the properties for that protocol.

More interestingly, we found that three of the nine combinations are not used in any of the existing protocols, and one such combination (report-agreed + server-matching), which we use S-DH to denote, provides an attractive combination of properties.

To analyze these protocols, we conduct a systematic analysis of the design objectives of PCT protocols, identifying four classes of properties: (1) Privacy, (2) Utility, (3) Resiliency to malicious attacks, and (4) communication/computation efficiency. Within privacy, we identify four different adversaries that have different capabilities and four types of personal information that may be revealed to different parties. We analyze how the two design dimensions impact these properties, and how specific features of existing protocols impact them.

**Contributions.** To summarize, our paper has the following contributions. First, we propose a general framework to design and analyze contact tracing protocols, identifying two main design dimensions. We conduct a comprehensive analysis of different designs' properties with respect to four aspects: Privacy, Utility, Resiliency, and Efficiency. We collectively refer to these properties as PURE. We analyze the PURE properties within the context of contact tracing and systematically compare the strengths and weaknesses of existing protocols with respect to the PURE properties. Our analysis enables the research community, industry, and governmental organizations to better understand the system model for these protocols and design options.

**Roadmap.** The rest of this paper is organized as follows. Section 2 discusses the framework of proximity-based contact tracing protocols, and the nine categories. Section 3 discusses the PURE objectives. Section 4 evaluates how the two design choices affect the properties. Section 7 discusses in more details some of the existing PCT protocols, especially when they introduce features not captured in the category. We discuss related work in Section 8, and conclude in Section 9.

## 2 A FRAMEWORK FOR UNDERSTANDING PCT PROTOCOLS

In this section, we introduce a framework for designing and analyzing PCT protocols.

### 2.1 Protocol Participants

We consider the following parties that are involved in PCT protocols.

**Users.** Any person who runs an app that executes a PCT protocol on a device she carries is a user. When two users are physically adjacent (e.g., within 6 feet) for some period of time (e.g., at least 15 minutes), we call that an **encounter event**.

**Patients.** Any user who is tested positive for the pathogen becomes a patient. An encounter event between a patient and a user is called an **exposure event**. PCT protocols aim to discover such exposure events.

**Public Health Authority (PHA).** Most contact tracing systems need to have some party testing users for the pathogen. We call such a party the Public Health Authority (PHA). PHA needs to have Personal Identifying Information (PII) of anyone who undergoes testing.

**Contact Tracing Server (CTS).** We use Contact Tracing Server (CTS) to model the entity that provides the functionality of communicating with all users (both non-patients and patients) to enable the detection of exposure events.

We first consider the setting where there is a single entity that controls both PHA and CTS. In Section 6.2, we discuss different architecture options and their implications in terms of privacy and resiliency.

### 2.2 An Abstract PCT Protocol

A PCT protocol has the following steps.
 (0) **Setup.** Each user downloads the app, which may exchange information with a server, such as setting up an account (which may or may not be linked with identity information).
 (1) **Exchanging Beacons.** Each phone broadcasts randomly generated **beacons** and simultaneously listens for beacons sent by nearby smartphones. These beacons (also called ephemeral IDs, pseudonyms, chirps, etc. in different protocol descriptions) are generated randomly and should not be linkable to users. In all PCT protocols, such beacons change periodically to avoid being used to reconstruct movement traces.

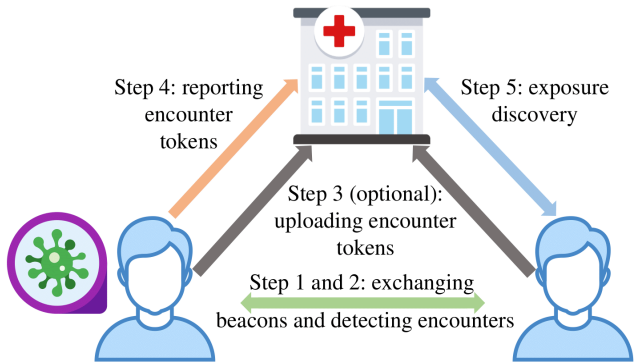

Fig. 1. Communications in a PCT protocol.

(2) **Detecting Encounters.** When a phone receives beacons, it can detect encounter events by estimating the distance from the source of the signals, and the duration of close encounter. When the distance and duration met conditions, the information associated with the encounter is stored. This information should be stored for a period of time depending on the nature of the diseases (e.g., 14 days to 3 months in different proposals).
(3) **(Optional) User Uploading Encounter Tokens.** In some protocols, periodically (e.g., several times a day), each phone may upload the encounter tokens to the CTS. In other protocols, nothing is uploaded by users at this stage.
(4) **Patient Reporting Encounter Tokens.** When a user becomes a patient, the server is notified, and the patient reports additional information to the server.
(5) **Exposure Discovery.** Some party computes the exposure events and the total exposure risk for each user, and inform the user when a user's exposure risk is high. How this is achieved is an important protocol design decision that we will explore below.

Figure 1 shows the interaction of the parties. We call the information uploaded by users (including patients and non-patients) **encounter tokens** (or **tokens** for short). In some protocols, encounter tokens are simply beacons; in others, tokens are computed from beacons.

## 2.3 Protocol Design Options

There are different ways to design a PCT protocol. We use the following two design dimensions to categorize PCT protocols.

**What do patients report?** Patients need to report encounter tokens so that exposure events can be detected, what those token contains is the first design dimension.

- **Report-Sent**. Patients report the beacons they sent out during the days they are likely to be infectious.
- **Report-Received**. Patients report the beacons they received during the days they are likely to be infectious.
- **Report-Agreed**. Patients report the tokens agreed between the two users in the encounter events that occurred during the days they are likely to be infectious.

**Who does the matching to discover exposure events?** The second design dimension is which party matches the tokens reported by the patients with those of users to identify exposure events. We consider three possibilities:

- **User-Matching.** Each user periodically downloads encounter tokens reported by new patients from the server, and checks against locally stored encounter information to identify exposure events.
- **Server-Matching.** In this approach, users need to upload encounter tokens periodically to the server. The server can thus match tokens reported by patients and those uploaded by the users to determine exposure risks for all users.
- **Interactive-Matching.** The server and each user jointly run some **interactive** protocol to perform matching. We have seen two major designs in how interactive-matching works.
 (1) **Querying.** Each user queries the server by uploading encounter tokens to the server and the server responds with exposure risks. The server does not store the user's tokens after answering the query. This reduces data storage and simplifies protection needs for the server, at the cost of increasing communication overhead. Each time the user queries, the user needs to upload all their relevant tokens, including those that have been uploaded in a previous query. This design can hide information regarding patients from the users, but does not hide information related to users' encounter events from the server.
 (2) **Secure Multiparty Computation (SMC).** The user and the server engage in a secure two-party protocol that reveals only specific results to the two parties. Typically, users learn their exposure risk, but nothing else. What the server learns depends on the design. For example, in one design, the server learns nothing except the number of users' tokens. In another, the server in addition learns the user's total exposure risk. In yet another design, the server also learns which patient's tokens led to a user' exposure.

|  | User-Matching | Interactive-Matching | Server-Matching |
|---|---|---|---|
| Report-Sent | DP3T [35], GAEN [12] | Epione [76] | |
| Report-Received | Clever Parrot [25] | | ROBERT [26] |
| Report-Agreed | Pronto-C2 [16] | DESIRE [27] | |

Table 1. Representative Protocols from the literature for the nine categories based on (1) what do patients report, and (2) which party performs matching.

Since each dimension has three possibilities, we have nine categories. Table 1 lists representative protocols that have been proposed in the literature. We have found protocols for 6 of the 9 categories.

## 2.4 A Quick Summary of Protocols in Nine Categories

We first give a brief review of the cryptographic primitives used in the protocols, and then give a summary of representative protocols in the nine categories. In addition to the six protocols in the literature, we introduce one protocol for each of the remaining three categories, as well as two basic versions of Sent-User and Received-User, which have slightly different properties from the protocols in thee categories from the literature.

Table 2 gives the key technical aspects of these protocols. In the next three sections, we focus on analyzing properties of these 11 protocols.

**Notations and Cryptographic Primitives.** We use $b, x$ to indicate values that a user computes and uses in the beacons she sent, and the *primed version* $b', x'$ to indicate values in a beacon she receives (computed by another user). We also note that in these protocols, a beacon is either of the form $H(s, t)$, which applies a pseudo-random function $H$ to a per-user secret seed $s$ and the current time $t$, or of the form $g^{H(s,t)}$, where $g$ is the generator of a group. We briefly review the security properties of $H$ and $g$ in the following.

| | Beacon Content | Patient Report | User Action |
|---|---|---|---|
| Sent-User, Basic | $b = H(s,t)$ | $b$ | Download patients beacons |
| Sent-User w/ daily seed (DP3T, GAEN) | $b = H(s_d,t)$ where $s_d = H(s,d)$ | $s_d$ | Download patient daily seeds $s_d$ |
| Sent-Interactive (Epione) | $b = H(s,t)$ | $b$ | Run PSI-CA |
| Sent-Server | $b = H(s,t)$ | $b$ | Upload received beacons |
| Received-User, Basic | $b = H(s,t)$ | $b'$ | Download beacons |
| Received-User (CleverParrot) | $g^x$ where $x = H(s,t)$ | $g^y \| g^{x'y}$ | Download patient tokens |
| Received-Interactive (RI-PSI) | $g^x$ where $x = H(s,t)$ | $g^{x'}$ | PSI |
| Received-Server (ROBERT et al.) | $b = H(s,t)$ | $b'$ | Upload $\{b\}$ |
| Agreed-User (Pronto-C2) | $g^x$ where $x = H(s,t)$ | $g^{xx'}$ | Download patients tokens |
| Agreed-Interactive (DESIRE) | $g^x$ where $x = H(s,t)$ | $g^{xx'}$ | Query w/ $H\left(g^{xx'} \| I(g^x < g^{x'})\right)$ |
| Agreed-Server (S-DH) | $g^x$ where $x = H(s,t)$ | $g^{xx'}$ | Upload $H\left(g^{xx'} \| I(g^x < g^{x'})\right)$ |

| | notations used in the table above, and their meanings | | |
|---|---|---|---|
| $s$ | a secret seed used by the user | $t$ | current time |
| $b$ | beacon sent by the user | $d$ | current day |
| $b'$ | beacon received by the user | $H$ | a pseudo-random function |
| $g^x$ | beacon sent by the user | $g$ | generator of a group |
| $g^{x'}$ | beacon received by the user | $1_{\{P\}}$ | indicator function that is 1 if $P$ is true and 0 otherwise |

Table 2. Summary of 11 protocol designs in the nine protocol categories.

A pseudo-random function (e.g., HMAC-SHA256 [77]) $H$ maps an input $X$ to an output $Y = H(X)$ deterministically, so that given $Y = H(X)$, it is hard to deduce input $X$, and given $Y = H(X)$, it is hard to find $X'$ so that $Y = H(X')$.

The generator $g$ is from groups for which computing discrete logarithms is difficult (e.g., cyclic groups and cyclic subgroups of elliptic curves over finite fields). Given such a group and its generator $g$, computing $g^x$ for any $x$ is efficient, but the reverse operation (computing $x$ from $g^x$) is hard. Moreover, knowing $g^x$ and $g^y$, one cannot compute $g^{xy}$.

**Sent-User.** We consider two protocols under this category. In the basic version, when a user becomes a patient, she reports to the server all the beacons she sent during the infectious period. Each user downloads from the server what the patients have reported, and then matches them with the received beacons she has stored to determine exposure events.

In the version used in DP3T [35] and the Google/Apple Exposure Notification (GAEN) framework [12], a daily seed $s_d$ is used so that all beacons used in one day can be derived from $s_d$. When a user becomes a patient, she reports one daily seed for each day during the infectious period. Each user downloads from the server these daily seeds, generates the patients' beacons using them, and then matches them with the received beacons she has stored to determine exposure events. This design reduces the amount of communications, but introduces privacy concerns.

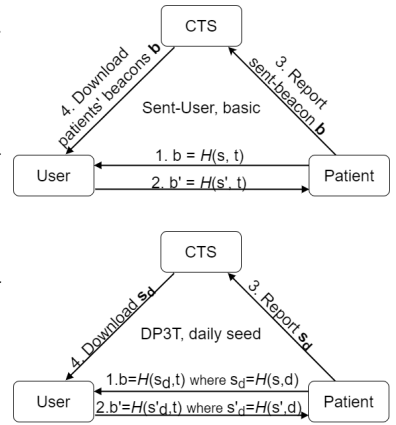

**Sent-Interactive.** Patients report what they sent. To determine exposure, each user runs a protocol with the server to match the received beacons she has stored with the set of all tokens reported by the new patients. We use Epione [76] as the example. Epione uses a Private-Set Intersection Cardinality (PSI-CA) protocol [33, 42] so that each user learns only how many beacons in her set are from patients. Compared with Sent-User, this offers better protection of patient privacy, but uses more computationally expensive cryptographic tools.

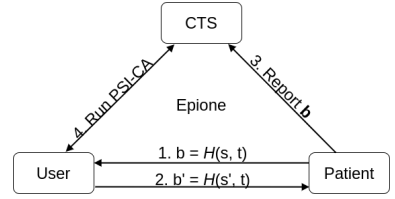

**Sent-Server.** A protocol in this category would have patients report the beacons they sent, and users uploading the beacons they heard so that the server can match them. Among existing protocols, none belongs to this category. This is likely because this design has a serious privacy flaw, which we will examine later.

**Received-User.**
We consider two protocols in this category. In the basic version, patients report the beacons they received, and each user downloads from the CTS the beacons reported by all patients, and sees whether these include the beacons she sent. For our analysis, we use a simplified version of the Clever Parrot [25] protocol that has similar properties. (See Section 7.4 for the actual Clever Parrot protocol.) In this version, beacons are of the form $g^x$. A patient reports, for each received beacon of the form $g^{x'}$, the token $g^y || (g^{x'})^y$, where $y$ is a newly generated random value. Only the user who sent $g^{x'}$ knows the value $x'$ and can tell that the token $g^y || (g^{x'})^y$ is computed by someone who has received $g^{x'}$. Other parties cannot link $g^{x'}$ and $g^y || (g^{x'})^y$ under the assumption that the Decision Diffie-Hellman (DDH) problem is hard.

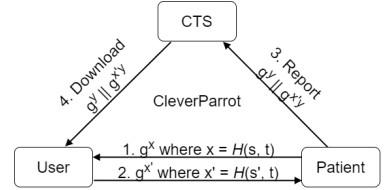

**Received-Interactive.**
In such a protocol, patients report the beacons they have received, and each user runs an interactive protocol with the server to determine exposure. In the literature, we have not seen a concrete protocol that falls into this category. To make the analysis concrete, we consider the following SMC-based protocol, which we call the RI-PSI protocol, as it uses a standard PSI protocol. In RI-PSI, beacons are of the form $g^x$. Each patient reports the beacons they received, e.g., $g^{x'}$, as their tokens. Each user uploads blinded versions of the beacons she sent by raising them to some secret $s$, ($g^{xs}$). Through interactions with the user, the server obtains patient-reported tokens, blinded using that user's secret $s$, and use them to match with what the user uploaded. The user learns the number of patients, how many beacons are received by each patient, and her exposure risk, but nothing else. The server learns only the amount of exposure for each user. We believe that this protocol has too high communication and computational cost to be practical. We introduce it here as an instantiation of the Received-Interactive category to enable our analysis.

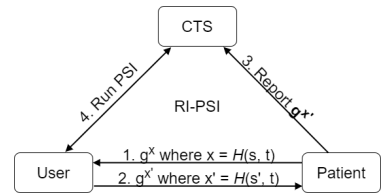

**Received-Server.** Patients report the beacons they received. To enable the server to match, the server needs to know which users send what beacons. In ROBERT [26], this is achieved by having the server generate the beacons used by the users. An alternative design is for users to report to the server which beacons they sent.

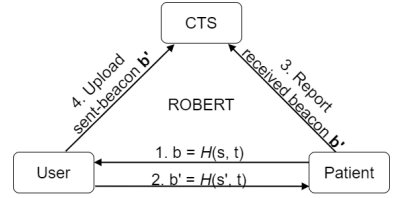

**Agreed-User.** Pronto-C2 [16] is an example of this category. Each user broadcasts beacons of the form $g^x$ and receives beacons of the form $g^{x'}$. Thus for each encounter event, both parties can compute tokens of the form $g^{xx'}$. When the user becomes a patient, she uploads these tokens. Each user downloads these patient tokens from the server and match them against stored tokens.

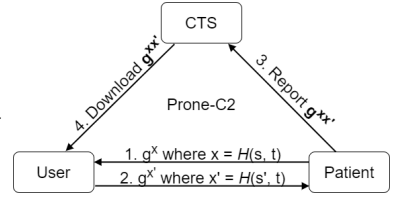

**Agreed-Interactive.**
DESIRE [27] also uses Diffie-Hellman-style tokens as in Pronto-C2, but users do not download patient tokens, and instead each user issues queries periodically. In each query, the user uploads a set of Diffie-Hellman tokens, and the server checks whether some of them are reported by the patients as well. To prevent the server from matching the tokens uploaded by two non-patient users, thereby learning that they were together, these tokens uploaded by users take the form

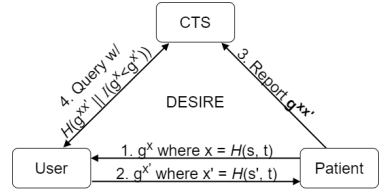

$H\left(g^{xx'} \;||\; 1_{\{g^x < g^{x'}\}}\right)$, where $1_{\{P\}}$ is the indicator function that has value 1 when $P$ is true, and 0 otherwise. This ensures that for the same encounter event, the two non-patient users would upload two tokens that are non-linkable by the server. When a user becomes a patient, it would report $g^{xx'}$, which can be matched with tokens of the form $H\left(g^{xx'} \;||\; 1_{\{g^x < g^{x'}\}}\right)$. In DESIRE, the server is not supposed to store each user's query content after answering it.

**Agreed-Server: the S-DH protocol.** We found no protocol in this category, although DESIRE comes close. DESIRE uses the query-approach of interaction. We propose a small change to DESIRE. Users periodically upload Diffie-Hellman tokens of the form $H\left(g^{xx'} \;||\; 1_{\{g^x < g^{x'}\}}\right)$ for their encounters, which the server stores. When a user becomes a patient, it would report $g^{xx'}$. The server uses this to compute encounter tokens and match with user tokens. We call this the S-DH protocol.

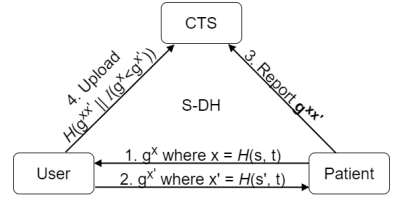

## 3 DESIRED PROPERTIES FOR PCT PROTOCOLS

We now identify the adversary model and the desired properties for PCT protocols. More specifically, we focus on four classes of properties: **Privacy**, **Utility** (accuracy and situation awareness), **Resiliency** (to malicious attacks), and communication & computation **Efficiency**. These four aspects (abbreviated as **PURE**) provide a comprehensive framework for analyzing and comparing contact tracing protocols.

## 3.1 Adversary Model

Several parties are involved in PCT protocols. As they play different roles in the protocols, their malicious activities can cause different effects.

**User-side Adversaries.** The user of a PCT system may be malicious or not fully cooperating. We consider four user-side adversaries with increasing capabilities.

- **Basic User.** A basic-user adversary uses the provided app to interact with the PCT system. Such a user tries to infer disease status about other users whom they have regular contact with, and may perform other malicious/non-cooperating activities.
- **Advanced User.** Some malicious user may be technically savvy. A PCT app can hide certain information from the user and perform validations on the user's actions to help prevent some attacks. However, relying completely on the app's trustworthiness for ensuring security/privacy properties is problematic, since the device is under physical control of the user. A malicious user may modify the app on the user's device to gain additional information, or download another app that speaks the same protocol, but records extra information and performs other actions that deviate from the specified behavior. Such an adversary can aim to compromise the privacy of other users, and/or to disrupt the PCT system.
- **User+Psv.** A malicious user can conduct surveillance by setting up devices (e.g. BLE sniffers) at various places to listen and record any beacons they hear. When these devices do not send messages, we use **User+Psv** (**Psv** for Passive SurVeillance) to denote such an adversary.
- **User+Asv.** This denotes a malicious user who deploys devices that both broadcast and receive beacons to engage in PCT protocols; here **Asv** is for for Active SurVeillance. The effect of active surveillance can also be achieved by multiple malicious users who collude and share the information with each other.

**Server-side Adversaries.** Here we assume that one adversary is in control of the server functionalities. Against a malicious server, we consider only privacy properties. No current protocol can prevent a malicious server from hindering the discovery of exposure events or creating false exposure events. Defending against them is likely to be prohibitively computational expensive. We consider three server-side adversaries with increasing capabilities.

- **Server-alone.** A malicious server can record additional information and modify its communication to attempt to learn more information about users (including patients).
- **Server+Psv.** A malicious server can also conduct surveillance by setting up eavesdropping devices at different places that aim to compromise users' privacy. Such an adversary can correlate information the server sees with that obtained through the surveillance.
- **Server+Asv.** This denotes a malicious server that deploy devices that broadcast and receive beacons to engage in protocols. This is the most powerful adversary against privacy.

**Discussions on Surveillance.** The scale of the surveillance depends on how many devices are used in the surveillance. Larger scale attacks require higher cost. While both server-side and user-side adversaries can perform surveillance, there are some differences. There is only one entity that controls the server. The entity is generally in a position of public trust and possesses a lot of resources. On the other hand, there are many malicious entities who control different sets of users. Therefore, for Server+Surveillance adversaries, we are more concerned with large-scale surveillance that covers a significant proportion of the whole user population. For User+Surveillance adversaries, however, we are more concerned with each adversary conducts surveillance targeting a certain geographical area, e.g., the campus of a corporation or government agency.

| **Adversary Model** | | |
|---|---|---|
| *User* | *Basic User* | Interact with the PCT system through the official PCT app |
| | *Advanced User* | Tampers with the PCT app (or download another one) |
| | *User+Psv* | Advanced user, also conducts passive surveillance |
| | *User+Asv* | Advanced user, also conducts active surveillance (complete protocol engagement) |
| *Server* | *Server-alone* | Use server-side data, i.e., what server receives from patients and non-patient users |
| | *Server+Psv* | Combine server-side data with data from passive surveillance |
| | *Server+Asv* | Combine server-side data with data from active surveillance |

Table 3. Adversary Model Summary.

While active surveillance is stronger than passive surveillance, engaging in active surveillance is more likely to be discovered, because active surveillance requires constant broadcasting. Passive surveillance would be harder to detect, since it can be conducted by well-hidden passive devices. We believe the fact that active surveillance is easier to be detected can provide some deterrence for a would be Server+Asv adversary. On the other hand, since there are many entities who can become user-side adversaries, and they may not be in positions of public trust, User+Asv adversaries are more likely to occur, although likely impacting a smaller set of users.

### 3.2 Privacy Objectives

In addition to the adversary model, there are two other dimensions to consider when evaluating privacy properties.

- **Information type.** Several types of private information can be leaked via PCT protocols. Through our analysis of the proposed PCT protocols and PCT processes, we identify four types of information: (1) ***identity of patients***, (2) ***exposure status***, (3) ***movement traces***, and (4) ***interaction relationship***. These will be discussed in detail below.
- **Information subject.** In most PCT protocols, patients need to send information that is different from other users. Thus, patients may have a higher level of privacy leakage than the users as a whole.

In summary, we consider two kinds of targets (all users and patients only), four types of information, and seven types of adversaries (four user-side and three server-side). There are $2 \times 4 \times 7 = 56$ combinations; however, not all are interesting. Some blatant leakages are prevented by all PCT protocols that have been proposed, whereas some leakages are unavoidable in order for contact tracing to happen. In the following, we identify the main privacy concerns by using the type of information as the top-level grouping. For each type of information, we consider whether and how the information could be leaked to different adversaries.

**Patient Identity.** When considering whether patients' identities can be leaked, we focus on user-side adversaries, as the server needs to know patient identity information as part of its function. A patient's identity is known by the PHA, which performs testing and informs the patient's result. Also in all protocols a patient needs to inform the server about the patient status, thus the server learns that an account belongs to a patient. Patient identity needs to be protected from users. If a user knows that the only possible exposure she has had is with one particular person, then she can infer that the other user is a patient when she is notified of an exposure. Such inference is unavoidable, and we do not consider this as leakage of patient identity if the protocol does not leak information that results in additional inference.

The main threat of leaking patients' identities to users comes from enabling users to learn the exact time they had an exposure event with a patient. A user may remember (mentally or use other apps to record) where she was at a particular time and other people who were present, providing information on patients' identities. Exposure time can be hidden from a basic-user adversary by not providing detailed exposure information to the user in the app's user interface. However, this protection is insufficient against an advanced user adversary. Thus, for patient identity, our focus is on **whether a protocol leaks exposure time to an advanced user adversary**.

**Exposure Status.** Here the issue is whether an adversary learns the aggregated exposure risk for users. This includes: the number of significant exposure events a user has, how many different patients a user has been exposed to, how long a user has been exposed to each patient, and so on. No protocol enables user-side adversaries to learn exposure status regarding other users. The main question is whether **the server can learn exposure information of users**.

**Movement Traces.** One main class of sensitive information that could be revealed by contact tracing protocols is movement traces. In traditional contact tracing, movement traces of non-patient users are never leaked to anyone. Patients may reveal some movement traces when discussing their contacts with contact tracers, but the amount of leakage is still subject to the control of the patients, who can provide contacts without locations. In location-based contact tracing protocols, the server generally learns movement traces, which is a major drawback of such protocols. One advantage PCT protocols have over location-based protocols is that leakage of movement traces can be better controlled because relative proximity instead of absolute location is used. However, movement trace information can nonetheless leaked when the adversary conducts surveillance. One major goal in designing PCT protocols is **to minimize leakage of movement traces to adversaries who conduct surveillance**.

**Interaction Relationships.** One consequence of leaking movement traces is to leak interaction relationships, which can be inferred from knowledge of the frequency and duration of two users appearing at the same location and time. Furthermore, even if movement traces are not leaked, some protocols may reveal the frequency and duration of two users appearing together, although the actual locations are unknown. Here we focus on leakage of interaction relationships beyond that implied by movement traces.

We distinguish between interaction relationships in situations that involve exposure events (i.e., involve a patient) and situations that do not require any exposure event (i.e., no patient involved). In the first case, there are three kinds of relationships: patient-to-patient, patient-to-user, and user-to-user (information that two users are both present together with some patient). We note that in traditional contact tracing, patient-to-patient and patient-to-user interaction information is leaked to the contact tracer, whereas User-Matching protocols can hide it. On the other hand, interaction information among non-patient users in contexts that do not involve exposure is never revealed in traditional contact tracing. We believe that leakage of this information must be prevented.

## 3.3 Utility Objectives

Contact tracing protocols aim to: accurately detect exposure events, compute the aggregated risk to a user caused by these exposure events, and inform the users whose aggregated risk is above some threshold. For utility objectives, we focus on two kinds of properties. First, when there are exposure events, the system should detect them. Second, the server may want to learn certain information that is helpful in fighting a pandemic.

**Ensuring Timely Patient Participation.** This is in regard to: *what is the impact of a user who stops or delays participation when being tested positive*. This property is affected by what patients report and how they report it.

**User-Risk Awareness.** This means that the server is aware of the aggregated exposure risk of each user. This enables PHA to focus manual contact and monitoring efforts on users that have high risks. We point out that it is the same as the exposure status privacy objective discussed earlier. Depending on one's perspective, this can be viewed as either an undesirable privacy leakage or a desirable feature enabling the server to be more aware of the situation to control disease transmission better.

**Transmission-Amount Awareness.** This means that the server is aware of how many exposure events have resulted from one patient. This helps identify what are known as super-spreaders, which are of epidemiological interests. We note that this information is readily available to the contact tracers in traditional contact tracing. This also enables identifying malicious attacks that aim to create panic or disrupt the contact tracing system, which manifest as one patient causing exposure of an unusually large number of users.

## 3.4 Resiliency Objectives

By resiliency, our focus is on the ability to defend against attacks carried out by user-side adversaries and aiming to create false exposure events that did not actually occur. Such attacks aim at creating panic and thus undermine the trust in a contact tracing system. In Section 3.1, we listed four user-side adversaries. Here we consider the specific attacks each adversary can carry out.

**Basic User.** Here the adversary is a single user who uses a normal phone running an PCT app. The main attack is that such a user falsely claims to be a patient, aiming to create false exposure events to undermine trust in the system.

**Advanced User.** Here we consider a malicious user who possesses special devices that can communicate using the PCT protocol. We also assume that the user can act as a patient after carrying out the attack. More specifically, we consider three kinds of attacks.

- **Drive-by Eavesdropping.** An adversary walks/drives around while collecting beacons that are being broadcast by users and act as if these beacons are from nearby users even if the estimated distances are far away and/or the contact is of short duration. Essentially, the malicious user ignores the distance and duration constraints when determining whether an encounter event has occurred. For example, in Report-Received protocols, when the adversary user becomes patient, it will report all these beacons, causing these users to be considered as having been exposed, even if they have been in contact only from far away and/or for only a short period of time.
- **High-powered Broadcasting.** The adversary uses high-powered broadcasting that can reach very far, misleading the recipient into thinking that they are in close proximity of the source of the signal. The goal is to cause people who receive the beacons to believe that they have been exposed when the adversary becomes a patient. This attack can potentially impact a large number of users when used in a large event where many people are gathering.
- **High-powered Device.** The adversary uses a device that has both high-powered broadcasting capability and high-sensitivity receiving capability that can receive signals sent far away. In a large event, one such malicious user can pretend to be adjacent to many users.

**Colluding Users.** This is equivalent to User+Asv adversary considered in Section 3.1. We do not specifically consider User+Psv as its effect is covered is similar to other attacks covered here. Here we consider multiple malicious users colluding, each using a normal device but possibly modified software. Their goal is to create false exposure events.

- **Same-Beacon.** The colluding users use the same beacons, instead of generating random beacons independently. The colluding users also share the beacons received with each other. If any of the users controlled by the adversary is confirmed as a patient, users with close contact to any of the colluding users would be considered to be exposed.

- **Pooling.** This can be viewed as a variant of the Same-Beacon attack. However, instead of all colluding users using the same beacon streams, they each generate their own beacon streams, but share the secret information among themselves. Each user's device impersonates all colluding users at the same time, broadcasting beacons belonging to all colluding users. If one of the colluding users becomes a patient, users with close contact to any of the colluding users would considered to be exposed.
- **Tunneling.** Colluding devices create tunnels between benign users that are far apart. This can be achieved, for example, using a shared bulletin board. Each device in real time sends to the bulletin board the beacons it received locally, downloads the beacons uploaded by other devices, and then broadcasts these beacons locally. For example, with $N$ malicious devices in $N$ different locations carrying out the tunneling attack, all users that are close to any one of the $N$ devices would think that they are close to each other. This boosts the number of encounter events roughly by a factor of $N$.
- **Forwarding.** A simpler form of tunneling attack that is easier to carry out is one-directional tunneling, which we call forwarding. For example, the adversary can forward beacons from many places to one target region, or forward beacons received from a target region to many other places, aiming to cause a large number of users in the target region to think that they have been exposed in near future.

## 3.5 Communication and Computation Efficiency

For communication efficiency, our focus is on the amount of communication between each user and the server. While each new patient needs to report some information to the server, in all protocols, the amount of communication for each new patient is proportional to the number of interactions the patient has, which seems reasonable. However, the amount of communication for each user can vary significantly, depending on the protocol design. The computation cost is mainly due to matching operations performed by the server or the clients. In most protocols, computation complexity is linear or linearithmic in the amount of communication; however, some protocols require expensive cryptographic operations.

**Resource Exhaustion Attacks.** In some protocols, the amount of communications from each patient is relatively fixed (e.g., Report-Sent protocols). In other protocols, the amount of communications depends upon the amount of encounters a patient has. A malicious patient can thus report a lot of data that consume communication and computation resources for servers and users, even if they cannot create false exposure events.

## 4 ANALYSIS OF THE PURE PROPERTIES

Having identified the main design choices in Section 2 and the desired properties in Section 3, we now analyze how the design choices affect the properties. More specifically, we analyze the PURE properties of the 11 protocols given in Table 2, and the results are shown in three tables. Table 4 shows the privacy properties, Table 5 shows the utility and resiliency properties, and Table 6 shows the communication and computation efficiency analysis. The rest of this section provides justifications and explanations for these tables.

Note that Tables 5, 6, and 6 together have about two dozen columns, representing that many properties. In order to present the analysis of about two dozen properties for 11 protocols succinctly, the analysis here is necessarily dense, and may be challenging to follow. The paper is structured so that one can return to this section to better understand the analysis when needed. In Section 5, we identify what we consider to be the most important flaws of PCT protocols, and analyze the most widely used PCT protocols. In Section 7, we consider additional protocol features that we choose to abstract away in this section.

| | | | Privacy Objectives | | | | | | | | |
|---|---|---|---|---|---|---|---|---|---|---|---|
| | | | Movement Trace | | | | | Interaction Relationships | | | |
| | | | All Users | | Patients | | | Involving Exposure | | | No-Exp. |
| **Category** | Exposure Status | Patient Identity | Server + Psv | Server + Asv | User + Psv/Asv | Server + Psv | Server + Asv | Patient to Patient | Patient to User | User to User | User to User |
| *Sent-User Basic* | ● | ○ | ● | ● | ● | ○ | ○ | ● | ● | ● | ● |
| *Sent-User w/ Daily Seed* (DP3T, GAEN et al.) | ● | ○ | ● | ● | ○ | ○ | ○ | ● | ● | ● | ● |
| *Sent-Interactive* (Epione) | ● | ◐ | ● | ● | ● | ○ | ○ | ● | ● | ● | ● |
| *Sent-Server* | ○ | ● | ○ | ○ | ● | ○ | ○ | ○ | ○ | ○ | ○ |
| *Received-User Basic* | ● | ● | ● | ● | ● | ○ | ○ | ○ | ● | ● | ● |
| *Received-User* (CleverParrot) | ● | ○ | ● | ● | ● | ● | ○ | ○ | ● | ● | ● |
| *Received-Interactive* (RI-PSI) | ○ | ● | ● | ● | ● | ● | ○ | ○ | ● | ● | ● |
| *Received-Server* (ROBERT et al.) | ○ | ● | ○ | ○ | ● | ○ | ○ | ○ | ○ | ○ | ● |
| *Agreed-User* (Pronto-C2) | ● | ○ | ● | ● | ● | ● | ○ | ○ | ● | ● | ● |
| *Agreed-Interactive* (DESIRE) | ○ | ◐ | ● | ○ | ● | ● | ○ | ○ | ○ | ● | ● |
| *Agreed-Server* (S-DH) | ○ | ● | ● | ○ | ● | ● | ○ | ○ | ○ | ● | ● |

Table 4. Privacy objectives of PCT protocols categorized in groups. The mark indicates whether an attacker can potentially obtain the information (○) or not (●). We use ◐ for patient identity leakage in Interactive-Matching protocols because some information can be leaked, but less than in User-Matching protocols.

## 4.1 Privacy Properties

**Patient Identity Leakage.** As discussed in Section 3.2, the main issue is whether users learn the exact time that their exposure to patients occur. Against a basic-user adversary, the app could be designed to hide this information. Against an advanced-user adversary, this depends on which party conducts the matching. In User-Matching (i.e., Sent-User, Received-User, and Agreed-User) protocols, users know which beacon(s) led to exposure events, and thus learn the time and location in which the exposure occurred. This leakage potentially enables users to identify the identity of the patient.

Interactive-Matching protocols generally try to alleviate this problem by having each user query the server through a cryptographic protocol. The user provides a set of beacons or tokens, and after the protocol, the user learns only the number of exposure events involving the provided set, but not which specific beacon in her set leads to exposure. However, this mitigation may be only partial. In such protocols, a user can query the server multiple times, each time with a selected set of beacons or tokens. If querying with one set leads to exposure warning, and querying with a subset of that does not, then the user learns which beacons are matched with patients.

Server-Matching protocols generally do not have such leakage as the server gives users only information of aggregated exposure risk without any extra information.

**Exposure Status.** Here the issue is whether the server learns how much exposure (to patients) one user has. This is also determined by which party conducts matching. Here, Server-Matching leads to leakage, User-Matching avoids such leakage, and Interactive-Matching depends on what the server learns through the interaction.

**Movement Traces of All Users.** Leakage regarding movement traces is the most complicated privacy challenge, as the vulnerabilities depend on many factors. Here we first consider movement traces of all users. To distinguish from privacy threats specific to patients, we use the term non-patient users here. Note that everyone starts as a non-patient, and some become patients.

In PCT protocols, apps broadcast beacons that are changed regularly and frequently (e.g., every 10 to 20 minutes). When these beacons are generated and broadcast correctly, different beacons by the same user cannot be linked together without additional information. Furthermore, no protocol provides information regarding non-patient users to other users. Thus there is minimal risk for the user-side adversaries to learn movement traces of non-patient users.

Since the Server has no location information, a Server-alone adversary cannot reconstruct movement traces. Whether movement traces is leaked to Server+Psv and/or Server+Asv depends on which party performs matching.

In User-Matching protocols, users do not need to upload any information, thus movement traces of non-patient users are protected against all adversaries.

In Interactive-Matching protocols, this depends on how much the server can learn about user's information through the protocol. For example, Epione [76] uses a private-set intersection protocol so that the server learns only the number of beacons the user heard; therefore, the server cannot learn movement traces. In DESIRE [27], for each query, a user uploads a set of Diffie-Hellman-style tokens of the form $g^{xx'}$, and the server matches them against data from patients. If the server does not conduct active surveillance, these tokens have no semantic meaning to the server, since only the sender of either $g^x$ or $g^{x'}$ can compute or recognize $g^{xx'}$ (under the Decision Diffie-Hellman (DDH) assumption), and no movement information is leaked. If, however, the server conducts active surveillance, the server can broadcast beacons at various locations, compute tokens for received beacons, and store them together with location and time information. Then the server can check whether the tokens uploaded by users match with stored tokens, where any matching pinpoints the location and time of the user. This enables reconstructing movement traces for users within the active surveillance area.

Let us now consider Server-Matching protocols. In Sent-Server or Received-Server protocols, users have to upload either what they received or what they sent to enable matching. This enables Server+Psv to construct movement traces for all users. In Agreed-Server protocols, users upload agreed tokens (of the form $H(g^{xx'}||1_{\{g^x<g^{x'}\}})$). While Server+Psv can receive all the beacons, the server cannot link a beacon to a token. However, a Server+Asv adversary can broadcast beacons in the surveillance area, receive beacons, and compute agreed tokens to reconstruct movement traces.

**Movement Traces of Patients.** Since patients have to report something extra in any PCT protocol, and their information may be shared with users (in User-Matching protocols), their movement traces are more vulnerable than non-patient users.

Let us first consider threats from user-side adversaries, namely user+psv and user+asv. Threats exist only in User-Matching protocols, in which users download patient information. The threat is small in Received-User or Agreed-User protocols if the server mixes data from all patients together before sharing them with the users, in which case the adversary learns only that some patient appeared in certain time and place in the past, but cannot link these appearances together to form a movement trace. However, for Sent-User protocols that use the daily seed feature to reduce communication cost (namely DP3T and GAEN), the adversary can tell that different beacons belong to the same patient, and thus can construct limited movement traces of patients.

We now consider threats from the Server+Psv adversary, where the threats depend on what patients report. When patients report what they sent, Server+Psv would receive those beacons in surveillance, and can obtain movement traces. When patients report agreed tokens, Server+Psv cannot link the reported tokens to anything, and cannot construct traces. When patients report what they receive, this further depends on who does the matching. In Received-Server protocols, Server+Psv can construct movement traces. In Received-User or Received-Interactive protocols, the tokens can be randomized version of the received beacons (of the form $g^y||g^{x'y}$) so that only the sender of the beacons can verify (as in Clever Parrot). As Server+Psv only listens and does not send beacons, Server+Psv cannot obtain traces.

The Server+Asv adversary both send and receive beacons, and can reconstruct movement traces for patients no matter what they report. Another way to think about this is that Server+Asv is essentially the server also playing the role of users that appear in many places. Since the server and users jointly must be able to determine whether the patient has encounters with the users, Server+Asv can reconstruct patient movement traces.

**Interaction Relationships against User Adversaries.** When a malicious server or user is combined with (active or passive) surveillance, it may learn movement traces and infer interaction relationships. That is already reflected in the analysis above. Below we focus on whether the adversary learns the number of times that two users were in proximity to each other without conducting surveillance.

Against user-side adversaries, potential leakage cannot occur in Server-Matching or Interactive-Matching protocols, because users do not have direct access to information from patients or other users. We consider User-Matching protocols, where each user can download information reported by patients. In Sent-User protocols, the reported tokens are the beacons each patient sent, which do not leak interactions. In Agreed-User protocols, the reported tokens are meaningful only for the two parties involved in that encounter, and do not leak interaction information either. In Received-User protocols, if any intersection in the token sets reported by two patients indicate that they were in close proximity. This can be prevented by the server merging data from all patients and removing duplicates. Thus we conclude that *interaction relationships do not leak to user-side adversaries*, and we do not list this property in Table 4.

**Interaction Relationships against Server Adversaries in Settings with a Patient.** Against server-side adversaries, we first consider interactions in settings where a confirmed patient is present, and we consider three kinds of relationships: between two interacting users who become patients around the same time, between a patient and a non-patient user, and between two non-patient users who appear together with a patient.

Let us first consider Report-Sent protocols. When patients report what they sent, such beacons are only about themselves and reveal no interaction relationships. Let us consider potential leakage from the information users need to upload when the server performs matching. In Sent-Server protocols, users upload what they receive. When two users upload the same beacon, the server knows that they were together. We emphasize that in Sent-Server protocols, the server learns interaction relationship between any two non-patient users (even when there is no patient involved).

Let us then consider Report-Received protocols. Since patients report to the server the beacons they received, the server can learn interaction relationships between two patients (when they report the same beacon) regardless of which party performs matching. When the server performs the matching, the server also knows which beacons are sent by which users. This enables the server to learn interaction relationships between a patient and a user and between two non-patient users in any context that a patient is present. More specifically, when two users' tokens appear in the same patient's report, the server can infer that they were in the same place at the same time.

Finally we consider Report-Agreed protocols. When patients report Diffie-Hellman-style tokens, the server can learn interaction relationships between two patients if they report the same $g^{xx'}$ token, regardless which entity is performing the matching. However, in Server-Matching and in the interactive model implemented in DESIRE [27], the server can also learn patient-to-user interaction relationships, but not user-to-user interaction relationships. This is because two users involved in a contact event upload the double tokens of the form $H\left(g^{xx'} \mid\mid 1_{\{g^x < g^{x'}\}}\right)$, so that the server cannot match the two tokens. In User-Matching, the server cannot learn interaction relationships neither between users and patients nor between two users.

**Interaction Relationships against Server Adversaries in Non-Exposure Settings.** For interaction relationship involving interactions among non-patient users in contexts that do not involve exposure, the main threat is for the Sent-Server protocols, where the patients report what they sent, and the server does the matching. To enable the matching, users must upload what they received. If a beacon appears in both users' uploaded set, then the server knows that these two users were physically close at some point. The size of the intersection of the uploaded sets from two users shows the number of times they were together. A graph over all users, where each edge shows the number of times they were together, is an interesting social network graph. The server gains this information even when there is no patient.

### 4.2 Utility Properties

**User Risk Awareness and Transmission Amount Awareness.** This is largely determined by which party conducts the matching. When the server does the matching, these properties are achieved, whereas when the users do the matching, there properties are not. In Interactive-Matching protocols, this depends on the specific output the server gets from the interactive protocols.

### 4.3 Resiliency

For resiliency, our focus is on preventing malicious adversaries from creating many false exposure events using one or more of the attacks discussed in Section 3.4. We want PCT protocols to be resilient to as many of these attacks as possible. Some of these attacks can be defended against because of what the patients report. For other attacks, additional detection and mitigation mechanisms are needed.

**Resiliency against Advanced User Attacks.** All Report-Sent protocols are resilient to drive-by eavesdropping attacks, since a malicious user reports only the beacons they sent, and not the beacons they receive. Furthermore, since each user should be sending only one beacon during a certain time period, it is straightforward for the server to limit how many beacon a patient should be reporting. However, this design does not prevent other attacks. For example, an adversary can use high-powered broadcasting to make her beacons reach many users, and then report these beacons when becoming a patient.

All Report-Received protocols can defend against high-powered broadcasting attacks, because what non-patient users receive does not contribute to exposure events. However, this design does not prevent other attacks. For example, a drive-by eavesdropping attacker can receive many beacons and then report them when becoming a patient.

In Report-Agreed protocols, an exposure event requires both involved parties to agree that they have been in close encounter with each other. These protocols can defend against drive-by eavesdropping and high-powered broadcasting attacks, as these attacks involve only one-direction spurious communication. For example, under a drive-by eavesdropping attack, since the target users did not receive any beacon, they would not store/report any token for this attack, and there would not be any exposure events. However, these protocols remain vulnerable to high-powered device attacks, which essentially enable an adversary to appear to be a user in close encounter with many users at the same time.

| | Utility & Resiliency Objectives | | | | | | | | |
|---|---|---|---|---|---|---|---|---|---|
| | Utility | | Resiliency | | | | | | |
| | | | Advanced User Adversary | | | Colluding Users Adversary | | | |
| **Category** | User-Risk Awareness | Transmission-Amount Awareness | Drive-by Eavesdropping | High-powered Broadcasting | High-powered Device | Same Beacon | Pooling | Forwarding | Tunneling |
| *Sent-User Basic* | ○ | ○ | ● | ○ | ○ | ○ | ○ | ○ | ○ |
| *Sent-User w/ Daily Seed* (DP3T, GAEN et al.) | ○ | ○ | ● | ○ | ○ | ○ | ○ | ○ | ○ |
| *Sent-Interactive* (Epione) | ○ | ○ | ● | ○ | ○ | ○ | ○ | ○ | ○ |
| *Sent-Server* | ● | ● | ● | ○ | ◐ | ◐ | ◐ | ○ | ○ |
| *Received-User Basic* | ○ | ◐ | ◐ | ● | ◐ | ◐ | ◐ | ○ | ○ |
| *Received-User* (CleverParrot) | ○ | ◐ | ◐ | ● | ◐ | ◐ | ◐ | ○ | ○ |
| *Received-Interactive* (RI-PSI) | ● | ○ | ◐ | ● | ◐ | ◐ | ◐ | ○ | ○ |
| *Received-Server* (ROBERT et al.) | ● | ● | ◐ | ● | ◐ | ◐ | ◐ | ○ | ○ |
| *Agreed-User* (Pronto-C2) | ○ | ◐ | ● | ● | ◐ | ◐ | ◐ | ● | ○ |
| *Agreed-Interactive* (DESIRE) | ● | ● | ● | ● | ◐ | ◐ | ◐ | ● | ○ |
| *Agreed-Server* (S-DH) | ● | ● | ● | ● | ◐ | ◐ | ◐ | ● | ○ |

| Summary of Utility Objectives | |
|---|---|
| *User-Risk Awareness* | Server knows the total exposure risk one user has |
| *Transmission-Amount Awareness* | Server knows how many users are exposed to one patient |
| **Summary of Resiliency Attacks** | |
| *Drive-by Eavesdropping* | Passively collects beacons broadcast by others, and report them when becoming a patient |
| *High-powered Broadcasting* | Broadcast beacons to reach users in a wide area, then become a patient |
| *High-powered Device* | Combines eavesdropping with high-powered broadcasting, using a high-sensitivity device |
| *Same Beacon* | Colluding users send the same beacon stream and shared received beacons |
| *Pooling* | Multiple colluding users share all data, and each device pretends to be multiple users. |
| *Tunneling* | Creates two-way communication "tunnels" between far apart regions |
| *Forwarding* | Forwards beacons received from one region to be broadcast in another far away |

Table 5. Utility and Resiliency objectives of PCT protocols categorized in groups. For Utility, we indicate whether the protocol provides the feature (●) or not (○). We sue ◐ to represent the case in which the server have some awareness, but not complete. For Resiliency, we represent whether one protocol is robust against the attack (●) or vulnerable (○). We use ◐ to represent the case in which server-side rate limit defense can be used to detect/mitigate the attack.

**Super-patient and encounter augmentation.** Let us think about encounter relationships between users as a graph where each user is a node, and two nodes have an edge when there is an encounter event between the users represented by these nodes. The different resiliency attacks can be classified as trying to achieve one of two possible goals. The first attacks aim to create a supernode that is connected with many other users, so that when that user becomes a patient many users are exposed. The single advanced user attack and beacon-sharing attack fall into this category. The other attacks aim to augment the graph by making it much denser without creating any supernode. For example, the tunneling attack adds edges between benign nodes. We now discuss how these attacks can be dealt with.

**Server-side Rate Limit.** The server can try to take actions to detect and mitigate super-patient attacks. A patient who claims to be in proximity of too many users at one time can be viewed as suspicious. When PCT is integrated with human-based contact tracing, the threshold can be chosen to be lower, as one patient exposing multiple users indicates either fraudulence or a super-spreader event, both worthy of investigation. In summary, the server can enforce a rate limit on the number of exposure events caused by one patient. While a malicious attacker can try to stay below the rate limit to avoid detection, this limits the impact of the malicious attack, reducing the incentive to carry out an attack.

However, not all PCT protocols can accommodate server-side rate-limit, In Report-Received and Report-Agreed protocols, for each encounter event, the patient must report something (either a received beacon or an agreed token); thus, it is straightforward for the server to apply the rate limit. In Report-Sent protocols, whether the server can apply rate limit depends upon which party does the matching. In Sent-User and Sent-Interactive protocols, the server does not learn how many users are exposed by one patient; thus the server cannot carry out the rate-limit defense, making these protocols vulnerable to resiliency attacks such as high-powered broadcasting attacks. In Sent-Server protocols, the server can learn how many users are exposed by one patient, and can accordingly apply rate limit.

Another form of server-side rate limit can mitigate resource exhaustion attacks, where malicious users generate fake beacons and tokens. Although the tokens cannot create exposure events, they can cause the server to spend many resources. By setting a threshold and rejecting more beacons/tokens from a (non-patient or patient) user, the server can limit the amount of resource consumption due to one malicious user. The threshold can be higher than the one used for identifying suspicious cases to investigate. Instead of examining the number of resulting exposure events, it limits the size of each user's reported input. Thus this can be applied in all protocols.

**User-side Rate Limit.** Against attacks that do not rely on supernodes, server-side rate-limit is less effective. In such attacks, e.g., pooling and tunneling, one malicious device pretends to be multiple devices at the same time. This attack can be detected and mitigated by the PCT app performing some rate limit on the client-side. For example, one defense is an app that, upon detecting the beacons of too many devices to appear to be nearby, performs a combination of the following actions: suppress these events, alert the user, and with user's approval, report to the central authority for investigation. It appears that user-side rate limit can be applied on all protocols.

## 4.4 Communication and Computation Efficiency

In almost all protocols, computation is linearithmic in the amount of communication. We thus focus on communication cost, which is mostly determined by which party performs the matching. When the server performs the matching, each user only needs to upload tokens related to herself, and one user's amount of communication is proportional to one's interactions. However, when the user performs matching, each user needs to download tokens from all patients, and the amount of communication for each user is proportional to the number of new patients times each patient's interaction amount. Since interactive matching typically aims to prevent the server from learning users' information, each user's communication amount is often as large as in the User-Matching case. Interactive matching also may require more expensive computations. For each user, the number of agreed tokens should be similar to the number of beacons the user receives, since the user computes an agreed token for each received beacon. This is generally higher than the number of beacons a user sends, since a user may receive multiple beacons at any time.

An exception to this is Epione [76], which uses a specially tailored Private-Set-Interaction Cardinally protocol. Each user, after interacting with the server to blind the beacons she received, run a Private Information Retrieval (PIR) protocol for each (blinded) beacon, to see whether it exists in the set of all patient tokens. There are several PIR protocols [8, 11, 19, 23, 38] one can choose from. In all of them, the server's computation cost for each PIR is linear in the server's input set (all patient tokens). Since each user needs to run PIR many times (once for each beacon the user received), we believe that the computational cost for the server is prohibitive.

| | Efficiency Overhead (Communication/Computation) | | | | |
|---|---|---|---|---|---|
| | Daily Overhead | | | | One-time |
| **Category** | User upload comm. | User download comm. | User comp. | Server Additional comp. | Patient comm. |
| *Sent-User Basic* | none | $14 \cdot 144 \cdot P$ | $(14 \cdot s) \cdot (14 \cdot 144 \cdot P)$ | none | $14 \cdot 144$ |
| *Sent-User w/ Daily Seed* (DP3T, GAEN et al.) | none | $14 \cdot P$ | $(14 \cdot s) \cdot (14 \cdot 144 \cdot P)$ | none | 14 |
| *Sent-Interactive* (Epione) | $14 \cdot s$ | $(14 \cdot s) \cdot log(C)$ | $(14 \cdot s) \cdot log(C)$ | Very high, see text. | $14 \cdot 144$ |
| *Sent-Server* | $14 \cdot s$ | none | none | $(14 \cdot s) \cdot (14 \cdot 144) \cdot N$ | $14 \cdot 144$ |
| *Received-User Basic* | none | $C$ | $(14 \cdot 144) \cdot C$ | none | $14 \cdot s$ |
| *Received-User* (CleverParrot) | none | $C$ | $(14 \cdot 144) \cdot C$ | none | $14 \cdot s$ |
| *Received-Interactive* (RI-PSI) | $144 + C$ | $C$ | $C$ | $(14 \cdot 144) \cdot C \cdot N$ | $14 \cdot s$ |
| *Received-Server* (ROBERT et al.) | $14 \cdot 144$ | none | none | $(14 \cdot 144) \cdot C \cdot N$ | $14 \cdot s$ |
| *Agreed-User* (Pronto-C2) | none | $C$ | $(14 \cdot s) \cdot C$ | none | $14 \cdot s$ |
| *Agreed-Interactive* (DESIRE) | $14 \cdot s$ | none | none | $(14 \cdot s) \cdot C \cdot N$ | $14 \cdot s$ |
| *Agreed-Server* (S-DH) | $s$ | none | none | $(14 \cdot s) \cdot C \cdot N$ | $14 \cdot s$ |

Table 6. Comparison of daily communication and computational cost in the 11 protocols. We assume that infectious period is 14 days, and 144 beacons are used each day (changing every 10 minutes). $P$: average number of **new patients** per day, $s$: the average number of contacts a user has per day, $C$: number of new tokens due to new patients (i.e., $14 \cdot s \cdot P$), $N$: total number of user.

| | Design Flaws | | | | | | | |
|---|---|---|---|---|---|---|---|---|
| **Category** | DF #1 | DF #2a | DF #2b | DF #2c | DF #3 | DF #4 | DF #5a | DF #5b |
| *Sent-User* (DP3T, GAEN et al.) | | | | | ✗ | ✗ | ✗ | |
| *Sent-Interactive* (Epione) | | | | | ✓̸ | ✗ | | ✗ |
| *Sent-Server* | ✗ | ✗ | ✗ | ✗ | | | | |
| *Received-User* (Basic) | | | ✗ | | | | ✗ | |
| *Received-User* (CleverParrot) | | | | | ✗ | | ✗ | |
| *Received-Interactive* (RI-PSI) | | | ✗ | | | | ✗ | ✗ |
| *Received-Server* (ROBERT et al.) | | ✗ | ✗ | ✗ | | | | |
| *Agreed-User* (Pronto-C2) | | | | | ✗ | | ✗ | ✗ |
| *Agreed-Interactive* (DESIRE) | | | | | | | ✗ | |
| *Agreed-Server* (S-DH) | | | | ✗ | | | | |

Table 7. Design flaws summary table. The ✗ mark represents a category that is flawed for a specific flaw. For Sent-Interactive (Epione), we use the crossed check mark to represent that the category is flawed to DF #3 under specific conditions.

## 5 COMPARING THE COMMON PCT PROTOCOLS

The analysis presented in Section 4 intends to cover all the aspects systematically. In this section, we distill the most important insights we gain from the analysis.

## 5.1 Ranking the Design Flaws

We list what we consider to be the most critical privacy, resiliency, and efficiency Design Flaws (DF) in PCT protocols.

**DF #1. Leakage of Non-exposure Interactions.** This refers to the leakage of user-to-user interaction relationships in non-exposure settings to a malicious server that does not conduct any surveillance. Such leakage happens in the Sent-Server design, because each user needs to upload the beacons they heard to the server. We believe that this one weakness suffices to rule out this design. No proposed protocol uses this design.

**DF #2. Leakage of User Movement Traces.** We consider three types of leakages here:
- **DF #2a:** Leakage of movement traces of all users to the Server+Psv adversary.
- **DF #2b:** Leakage of movement traces of patients to the Server+Psv adversary.
- **DF #2c:** Leakage of movement traces of all users to the Server+Asv adversary.

We consider DF #2a to be the most serious, as it impacts all users and passive surveillance can be carried out stealthily. It also implies DF #2b and DF #2c. A PCT system with DF #2a provides the entity who runs the PCT service with a new surveillance capability. Other than Sent-Server protocols, Received-Server (to which ROBERT belongs) is the only category that has this weakness.

**DF #3. Leakage of Patient Identity to Users.** This refers to leakage of patient identity to users, which happens because users can identify the specific time that an exposure occurs. An important principle of traditional contact tracing is that, quoting from [2] (Italic used in the source):

> *The identity of the patient or other identifying information will not be revealed, alluded to, or confirmed by the contact tracer, even if explicitly asked by a contact.*

That is, users should learn no additional information besides whether they have had contact with some patients. All User-Matching protocols (including DP3T, GAEN, CleverParror, and Pronto-C2) have this flaw.

**DF #4. Resiliency Vulnerability with no Mitigation.** This refers to vulnerability to either the drive-by eavesdropping attack or the high-powered broadcasting attack with no ability to detect/mitigate such attacks. Sent-User and Sent-Interactive protocols have this flaw. An adversary can use high-powered broadcasting to reach many users. When the adversary becomes a patient, a large number of users will be notified that they have been exposed. Such an attack can be combined with pooling or forwarding attacks, where the adversary can broadcast multiple beacon streams at the same time. The adversary may do this when it cannot ensure that any single stream can be reported later as belonging to a patient. Since the patients report what they sent, which do not yield information on how many users may be exposed, and the server does not perform matching and does not know how many users are exposed to one patient, the server cannot deploy server-side rate limit to detect or mitigate the attack. The impact of the attack is a large number of fake exposure warnings, eventually undermining the trust in the system.

**DF #5. High Communication/Computation Cost.** We consider two forms of high cost.
- **DF #5a.** The communication of each non-patient user is linear in the number of new patients. As this number can be very large, the communication cost is significant and impacts all users.
- **DF #5b.** Relies on expensive cryptographic protocols that has never been used in large-scale deployment.

DF #5a affects all User-Matching protocols. As an evidence that this is a serious issue, all contact tracing apps that deploy User-Matching protocols such as GAEN or DP3T introduce other features to reduce the amount of communication, even though doing so hurts privacy. DF #5b includes Epione, which requires many executions of private information retrieval in each user query.

## 5.2 Sent-User Basic vs. Received-Server (ROBERT)

A lot of discussions on PCT revolve around comparing decentralized design with centralized design, where DP3T and GAEN are examples of the decentralized design, and ROBERT is an example of the centralized design. Here we compare them using our PURE framework. In summary, DP3T and GAEN have DF #3, #4, and #5a, and ROBERT has DF #2a.

**Privacy.** The main privacy advantage of the Sent-User approach is that it leaks no information about non-patient users. Its main privacy weakness is that it leaks patient identity to users through leakage of precise exposure times (DF #3).

Server-Matching protocols provide the best protection against leakage of patient identity, but offer less protection for non-patient movement traces information. The Received-Server design leaks non-patient movement traces to server+surveillance adversaries (DF #2a), and interaction relationships in situations that involve exposures.

**Utility.** Sent-User protocols do not enable user risk awareness or transmission amount awareness, meaning that the server does not know how many users are exposed by one patient, and how much risk each user has. Received-server protocols enable the server to learn this.

**Resiliency.** Sent-User protocols are vulnerable to high-powered broadcasting, and Received-Server protocols to drive-by eavesdropping. However, in Received-Server protocols, the server learns how many users are exposed by one patient; the server can then apply rate limiting to defend against many resiliency attacks. Sent-User protocols cannot deploy this defense and would be particularly vulnerable (DF #4).

**Efficiency.** In Sent-User basic protocols, each day a user needs to download $14 \cdot 144 \cdot P$ beacons, where $P$ is the average number of new patients per day. This dependency on the average number of new patients is a serious concern (DF #5). Several efforts were made to mitigate this, but they all introduce other issues. We discuss these in Section 7.1. In Received-Server protocols, the amount of communication for each user is linear to the number of beacons she sent, and that for each patient is linear to the number of encounters she has. These are reasonable.

## 5.3 Using Agreed Tokens

Our analysis shows that Report-Agreed protocols offer advantages compared to Report-Sent and Report-Received ones. Here we discuss this in more details.

**Privacy.** An important privacy advantage is that a token of the form $g^{xx'}$ is meaningful only to the two parties involved in generating the token (one knowing $x$ and the other knowing $x'$). This reduces information leakage in Server-Matching protocols (which have better utility and resiliency properties), making them more privacy-friendly. For example, while Sent-Server leaks user-to-user interactions in all settings (DF #1), Received-Server does so only in exposure settings, and Agreed-Server does not leak. Furthermore, both Sent-Server and Received-Server leak movement traces of all users to Server+Psv (DF #2a), whereas Agreed-Server does not. For an server adversary to learn movement traces, it must conduct active surveillance.

**Utility.** When PCT uses an underlying communication channel that is not fully duplex, then one device may miss the beacons sent by another device when they happen to be listening at the same time. Even when randomization is used to deal with this, there is always some probability that one device fails to receive another one's broadcast. Even when full duplex communication channel is used, the probability that one device failing to receive beacon from another is non-zero due to poor transmission or other signals. In Report-Agreed protocols, both parties need to receive the beacons from the other party in order to identify the encounter event; therefore the failure probability is about twice that under Report-Sent or Report-Agreed protocols, for which only one-direction communication needs to succeed. When the failure probability is low, this increase in failure probability has a small impact.

**Resiliency.** In terms of resiliency, what patients report has a critical impact on each protocol's robustness. Report-Received protocols are vulnerable to an eavesdropping attack, whereas Report-Sent protocols are vulnerable to broadcasting attack. In Report-Agreed protocols, a contact token is valid only when both users participate in the exchange process. This approach provides better resiliency. However, other resiliency attacks can still be effective (see Table 4.3). Additional defensive solutions (e.g., rate-limit approaches) are necessary to mitigate these attacks.

**Efficiency.** In Report-Agreed protocols, when a (patient or non-patient) user needs to upload information, the amount of communication is linear in the number of encounter events that user experienced. This is similar to Report-Received protocols. When the server performs matching, the server needs to perform one hashing for each token before matching. This is also acceptable, as hashing is inexpensive.

## 5.4 Comparing three Report-Agreed protocols

We now discuss the main differences among Agreed-User, Agreed-Interactive, and Agreed-Server protocols. In summary, Agreed-User has DF #3 and DF #5a, Agreed-Interactive (DESIRE version) has DF #2c and partially DF #3, and Agreed-Server has DF #2c.

**Privacy.** All User-Matching protocols suffer from DF #3 (leakage of patient identity), and Agreed-User is no exception. Agreed-Interactive design partially avoids DF #3; however, it has DF 2c (leakage of user movement traces to Server+Asv). Agreed-Server fully avoids DF #3, but has DF #2c.

**Utility.** In Agreed-Server and Agreed-Interactive (in the DESIRE version), the server has situation awareness. In Agreed-User, the server does not.

**Resiliency.** In terms of resiliency properties, as the patients report agreed tokens, all the three agreed categories achieve the same result. Since the number of exposure events is bounded by the number of encounter tokens one patient reports, the server can perform rate limiting as a defense.

**Efficiency.** All User-Matching protocols suffer from DF #5a (high user communication cost), and Agreed-User is no exception. In Agreed-Server and Agreed-Interactive protocols, each user's communication cost does not depend on the total number of patients, and only on the number of encounter events that user has. Agreed-Interactive has higher communication cost than Agreed-Server, because for each query the user must upload all beacons.

## 5.5 Conclusions from the Analysis

The most importance conclusions from our analysis are the following.

- One can choose at most two properties among: (a) Strong user privacy, (b) Strong patient privacy, and (c) Low communication cost. More specifically, User-Matching provides (a), Interactive-Matching can provide (a) and (b), and Server-Matching provides (b) and (c).
- Reporting agreed tokens offers better privacy and resiliency properties than other designs, no matter which party performs matching. Thus the Agree-Server design mitigates the most serious user-privacy concerns in Sent-Agreed and Received-Agreed protocols, while preserving (b) and (c) above, making it an attractive design.

## 6 OTHER ISSUES IN PCT PROTOCOLS

We now discuss other features and issues for PCT protocols that are somewhat orthogonal to the two dimensions of what patients report, and who performs the matching.

## 6.1 Other Useful Features for PCT Protocols

**Exposure Aggregation.** The US CDC defines a close contact as any individual within 6 feet of an infected person for at least 15 minutes.[2] When beacons change every 10 or 20 minutes (as suggested in most proposals), in any continuous 15 minutes, one user may use multiple beacons (up to two if beacons change every 20 minutes, and up to three if beacons change every 10 minutes). When a user A is close to a user B for 15 minutes, user A may see two beacons from B, each for less than 8 minutes. It is thus necessary for PCT protocols to enable aggregation over multiple beacons.

Another similar challenge, which is rarely discussed in the literature, is how to deal with **intermittent contacts**. That is, two users may be together for several minutes, then move away, and then be back together for several minutes. This can happen in many situations, e.g., in outdoor gathering or public transportation where people move around, colleagues who meet each other for a short duration multiple times each day (in elevators, public space, etc.). The threshold of 15 minutes can be reached through the aggregation of several sessions of contact.

One way to deal with these challenges is to set a shorter time threshold for what is considered to be an exposure event in contact tracing. For instance, one can define exposure events to be at least 2 minutes long. Then the exposure risk from multiple events can be aggregated to compute an aggregated exposure risk. Only very short exposure events (e.g., shorter than 2 minutes) will be missed. While most proposals do not explicitly discuss how to support the aggregation of multiple short exposure events (such as intermittent contacts), it is fairly straightforward to add this feature to most existing PCT protocols. Adding this feature, however, would increase the number of exposure events, which adds to the communication cost.

**Filtering Events with Household Members.** For many people, the vast majority of encounter events are between household members who will inform each other when one becomes a patient. For protocols where communication costs depend on the number of encounter events, reporting encounter events between household members unnecessarily increases the communication cost. One possible design is to enable two users who mutually agree to inform each other in the case of one becoming patient to not store or report encounter events between them. This can be achieved by having two household members exchanging the secret seeds used to generate beacons, e.g., by scanning QR codes. Depending on the protocol design, not uploading such information may improve privacy, as it hides information regarding interactions within household members.

## 6.2 Server Architecture Considerations

Recall from Section 2 that we consider both a Public Health Authority (PHA) and a Contact Tracing Server (CTS). So far, we have assumed that one entity controls both. We now consider additional architectures. We refer to theses architectures using S0, S1, S2, S3, SN, where the number is related to the number of entities acting with some server capacity in the system.

- **S0.** In this architecture, PHA is not involved in the contact tracing system. There are apps working in this mode (e.g, [30, 59]). Without involvement of PHA, each user can self-identify as a patient.
- **S1.** One single entity (such as the government) provides both the PHA and the CTS. Since PHA needs to have Personal Identifying Information (PII) of anyone who undergoes testing, this means that the CTS has the identity information of (at least) a significant fraction of the population. In some systems that use the S1 architecture, such as Bluetrace [17], all users provide PII when they register; thus the CTS knows everyone's PII.
- **S2.** The PHA and the CTS are run by two separate entities. PHA has PII. Users have accounts on the CTS and communicate with the CTS; however, these accounts do not need to be linked

- **S3.** There is a third server (in addition to PHA and CTS) that provides some services. For example, in TraceSecure [18], a third server is used to provide anonymous communication between the users and the CTS, so that each user's data is provided to the CTS under a pseudo-nym. For another example, in Epione [76], an auxiliary server is used in the PSI-CA protocol. Compared to S1, the functionalities of the server is split into three servers so that when they do not collude, each party learns more limited information regarding users.
- **SN.** Note that the main role of the CTS is to enable (indirect) communications between patients and non-patient users so that their data can be matched to determine exposure. Such a role can be played by a peer-to-peer system. Without a central server, one cannot have Server-Matching or Interactive-Matching protocols. Thus this architecture is only compatible with User-Matching protocols. Such a design is likely to incur higher communication cost, and is potentially easier to be disrupted by a small number of malicious users.

**Architecture's Impact on Patient Status Authentication.** In Section 4.3, we discussed defenses against an adversary who tries to create large amount of false exposure events. With such defenses, we can limit the attack impact caused by each individual patient account. To achieve sufficient adversarial impact, an adversary may try to obtain access to other accounts (possibly by paying other users). We now consider the effects of such attacks.

*False Claim of Patient Status.* Here the adversary convinces other users (possibly with payment) to falsely claim to be a patient. This attack is effective only in the S0 architecture, where there is no PHA and anyone can claim to be a patient. To defend against this attack, we need the involvement of the PHA. One approach is for the PHA to provide users with an authorization code when they are diagnosed positive. The authorization code enables one patient to report data. This approach, however, is vulnerable to the attacks in the following.

*Transferring Patient Status.* The adversary pays patients for their authorization codes, and ask them to report data from accounts they directly control. A defense against such an attack is to require each user commit to a CTS account at testing time. Once a user is tested positive, only the committed account can be used to report data. For example, the user can be required to provide a commitment $H(\text{acc}||r)$ at testing time, where acc identifies the user's account, and $r$ is a random string. The authorization code should be tied to $H(\text{acc}||r)$, which can be achieved either by cryptographic means (such as PHA digitally signing $H(\text{acc}||r)$), or simply by the PHA sending to the CTS pairs of authorization codes and such commitments. This approach, however, remains vulnerable to the following attack.

*False Patient Data.* The adversary pays patients for reporting data the adversary gives them. This attack requires higher effort from the patients, as they may need to install a new app to report the data. Thus it is more expensive for the adversary to carry out. To defend against this attack, a PCT protocol can require that at testing time users commit to the data they would need to report once they become patients. The Agreed-Server protocol (S-DH) already does this. In the S-DH protocol, as a user, one uploads tokens of the form $H(g^{xx'}||b)$, and after becoming a patient, one reports $g^{xx'}$, which can be verified to match the tokens uploaded earlier. In other protocols, one could require that at testing time, a user uploads an encrypted version of the data she would need to report after becoming the patient. If the user is tested positive, she only needs to provide the decryption key. While it is still possible for the adversary to pay users before they test, so that they commit to data provided by the adversary, the attack is less cost effective, as the testing may come back negative. The adversary would then need to pay more in order to obtain the same number of accounts to report false information.

*Summary on Patient Status Authentication.* From the above discussions, we can see that there are differences in terms of patient authentication and privacy whether we assume S0 (no PHA), S1 (one party controlling both PHA and CTS), and S2. We do not see S3 or SN resulting in significant differences from S2.

**Ensuring Timely Patient Participation.** A further modification of the above design is to require that each user, before being tested for the pathogen, uploads encrypted version of the data she would need to report after becoming patient to one of PHA and CTS, and hand the decryption key to the other server before testing. When the test result is positive, PHA shares the data with CTS, enabling the decryption of the data.

This design can deal with the situation that *a user who tests positive chooses not to enter this information in the app, or enters the information after a long delay.* For example, a patient may simply choose to stop using the app after tested positive. This can happen even if the user is not malicious, as a person who is tested positive may be too overwhelmed to enter this information in a timely fashion. This design ensures that there is minimal delay from the time testing result comes out to the time that other exposed users can be notified, as the patient does not need to involved.

When one cannot trust the CTS and PHA not to collude, this approach is equivalent to each user reporting patient data to the server before testing. The impact of doing this is that privacy concerns that were only applicable to patients, are now more broadly applicable to anyone who is tested.

**Architecture's Enhancements for Privacy.** One can use multiple servers to compartmentalize the amount of information each individual server knows, thus reducing privacy leakage to each individual server.

*Improving Patient Privacy.* Recall that in many PCT protocols there are some leakages of patient movement traces to the CTS especially with surveillance. Hence, preventing the CTS from accessing users' PII would provide better privacy. As the PHA needs to test and verify patients' status, it must obtain patients' PII. A multi-server design can break the link between PII and contact tracing data by distributing them across different entities. Epione [76] and ContraCorona [21] propose this type of solution, where patients tokens are encrypted under CTS's public key. Once the PHA confirms the patient's status, it forwards the encrypted information to the CTS, which possesses the private key to decrypt the tokens. The CTS can then mixes the tokens from different patients and forwards them to another server that computes the matching. By receiving tokens from multiple patients mixed together, this second server cannot link the tokens to specific patients.

*Improving User Privacy.* The users' privacy can also be improved by introducing an additional entity to break the linkability between users' identities and tokens. In TraceSecure [18], the authors propose a design that uses three servers. Server 1 behaves as PHA that collects and mixes patients' tokens (received beacons in this case) and sends the tokens to Server 3. Server 2 collects users' information and associates them with pseudonyms before sending the information to Server 3. In TraceSecure, users need to upload the seeds from which they generate their beacons. While Server 3 has tokens from patients and users, they are only associated with pseudonyms. After performing the matching, Server 3 sends the result to Server 2, who forward them to users.

### 6.3 Limitations

PCT protocols have limitations and face challenges that are independent from the protocol and architecture design.

**Effect of the Metcalfe's Law.** Effective contact tracing requires that a large percentage of the population use it, and that people who interact with each other use the same, or at least interoperable systems. One fact about PCT technology is that an exposure event can be detected only when both parties in the event use the technology. Therefore, if the fraction of users using the technology is $p$, the probability that a particular exposure event is detected is approximately $p^2$. This is similar to the Metcalfe's law [57], which states that the effect of a telecommunications network is proportional to the square of the number of compatible communicating devices in the system.

Note that this $p^2$ is a rough estimation. On the one hand, $p^2$ may be an underestimation. For a hypothetical example, suppose that the population is divided into two groups each having half of the population, and interactions happen only within each group. If the deployment ratio in one group is $2p$, and that in the other is 0, then one can expect to capture $2p^2$ (instead of $p^2$) of all interactions. On the other hand, $p^2$ may be an overestimation. When the usage of PCT is voluntary, there is likely a bias in the group who voluntarily adopt the technology. People who are more conscientious about the pandemic are more likely to use it. These people are also more likely to deploy other mechanisms (such as social distancing and wearing facial covering) to prevent transmission. People who do not deploy other prevention mechanisms and hence being more likely to be involved in a transmission are also less likely to download the contact tracing apps. A deployment ratio of $p$ may be able to capture less than $p^2$ exposure events.

These caveats notwithstanding, we believe that $p^2$ provide a reasonable approximate estimation. The implication of this fact is that PCT technology can play a significant role only when it has a very high deployment rate. In order to catch half of all exposure events, we need a deployment rate of over 70%. Note that it is estimated that 81% of the US population uses smartphones [61, 65].

For similar reasons, fragmentation of PCT systems also reduces the effectiveness of contact tracing. If 70% of the population all use some PCT technology, but they use apps that do not interact with each other, then contacts between users who use different apps cannot be captured by the system. Increasingly, new PCT systems are built based on the GAEN framework, which help with inter-operability.

**Accuracy Challenges.** There are some other inherent limitations when using mobile apps to detect exposure events. Even for users who have a phone and use a PCT app, they may not carry their phones with them all the time. Moreover, distance between phones may not be estimated accurately. BLE-based PCT techniques uses signal strength to estimate the distances [45, 53, 55]. However, the distance between two phones may not accurately reflect the distance between the persons carrying them, e.g., the phone may be in backpacks or pockets. Distances between two persons may not fully reflect the actual risk. For example, the likelihood of transmission between two people sitting together eating and talking is very different between two people sitting together in public transportation with face-masks on.

## 7 ADDITIONAL VARIANTS OF PCT PROTOCOLS

In this section we discuss proposed PCT protocols, grouped by their categories. If the PURE properties are already fully determined by the category, as discussed in Section 4, we do not repeat them here. We focus on the situation when a protocol has interesting additions on top of the base design in a category.

### 7.1 Sent-User Protocols

Several prominent protocols belong to this category, such as the Google/Apple Exposure Notification (GAEN) protocol [12, 41], DP3T [35, 68], which comes in three versions, DP3T Low Cost, DP3T Unlinkable and DP3T Hybrid. Other protocols in this category include ReBabbler [25], TCN [75], East Coast PACT [28], West Coast PACT [69] and Stop Corona[15].

To deal with DF #3 (leakage of patient identity) and DF #5a (high communication cost), protocols in this category introduce additional features.

**Daily Seeds.** To deal with high communication cost, most protocols in this category employ daily seeds. Thus, users generate all the seeds of one day from the same seed and patients upload only the seeds to enable other users to generate the beacons and perform matching. For each day a patient needs to upload only one seed, instead of 144 beacons. While this approach reduces users' communication significantly (e.g., by a factor of 144), it exacerbates the leakage of patient identity, as all beacons generated from one seed are linked together. It also potentially leaks patient movement trace information to parties conducting surveillance. For example, if an adversary conducts surveillance on a college campus, when daily seeds are used, the adversary can create daily movement traces of patients.

One could also choose to make each seed valid for shorter periods of time than one day, e.g., 2 to 4 hours. This alleviates the movement leakage somewhat as each trace is shorter, but increases communication cost. Under this design, patients do not receive privacy benefits from frequent changes of beacons, since these beacons can be linked.

**Probabilistic Matching.** DP3T Unlikable [35] adds another twist to address the privacy problem caused by daily seeds. The server uses the seeds to generate all patients' beacons, and then use a Cuckoo filter [36] as a probabilistic representation of a set of beacons. Users download the filters and perform matching. Since users do not see the seeds, they cannot link together one patient's beacons. This has higher communication cost than using daily seeds, but lower cost than directly using beacons. However, it introduces false positives in matching, which may also be a usability concern, as warnings from a protocol that by design introduces false positives are more likely to be ignored, even if the false positive probability is low.

**Summary.** We believe that the need to introduce the above features, which sacrifices either privacy or accuracy, is a manifestation of the efficiency weakness. The root cause of is that each user's communication cost depends on the number of new patients, which is large. These features do not address patient identity leakage or resiliency concerns.

## 7.2 Sent-Interactive Protocols

We use Epione [76] as the representative of Sent-Interactive protocols. In Epione, patients upload what they sent, users learn their exposure risk by running a Private Set Intersection Cardinality (PSI-CA) protocol such that each user learns the cardinality of the intersection of the set of patient beacons possessed by the server, and the set of beacons the user has received. More specifically, the server blinds all patient beacons using $s$ (raising them to the $s$-th power). By interacting with the server, each user obtains a blinded version (also using $s$) of all beacons she has heard, without leaking information about the beacons to the server. Then the user runs a Private Information Retrieval (PIR) [5, 11, 24, 34, 38] protocol for each blinded beacon.

In Epione, the user performs a PIR for each beacon she has received. While it uses a PIR protocol whose cost is log of the size of the set asymptotically, the computational and communication cost of PIR remains fairly high. It seems unlikely that this protocol can be made practical for the purpose of contact tracing in near future.

## 7.3 Sent-Server Protocols

As mentioned in Section 2, we have not found any protocol that falls into this category. If the server does the matching when patients report what they sent, all users need to report what they received. When two users report the same beacon, it means that they were in close proximity of each other at the same time. The server can thus learn the frequency and duration of users in close proximity of each other, enabling the server to learn interaction relationships between non-patient users, without any surveillance attacks. In our opinion, this leakage makes such protocols unacceptable.

## 7.4 Received-User Protocols

An example of this category is CleverParrot [25], in which users broadcast beacons of the form $b = H(e)^s$, where $H$ is a cryptographic hash function, and $e$ is some measurement of the environment and could depend on the time and/or location and $s$ is the secret the user has. It is required that two devices in close proximity should compute the same $e$. Users store the pair $(b, e)$ for each received beacon $b$. (Earlier in the paper, we analyzed a simplified version of this protocol where everyone uses a public value $g$ as the base of exponentiation, instead of using an environment-dependent $H(e)$. This avoids the need for the PCT app to have location information.) A patient generates a random $y$ and reports to the server the value $(b^y, H(e)^y)$ for each stored pair $(b, e)$. Users can check their status by downloading patient-reported tokens and, for any token $(u = b^y, v = H(e)^y)$, checking whether $u$ equals $v^s$. A match indicates an exposure event. The use of $y$ randomizes the reported beacons, such that only the user who generates the beacon can recognize it. This is not a report-agreed scheme, as the token does not depend on what the patient sent, and a user needs only her secret value $s$ to match the token. In Clever Parrot, a user does not learn directly which of her beacons $b = H(e)^s$ results in matching with a patient token $(x = b^r, y = H(e)^r)$; the only thing the user learns is which $s$ value matches the token. However, one privacy issue, also pointed out in [25], is that a malicious user can change $s$ frequently to uniquely identify each contact and learn exposure time. The countermeasure suggested in [25] is to introduce a registration authority that provides "certificates" for all the $s$ values used by users. This, however, introduces another trusted party and significant overhead, and, as mentioned by the authors, certifications for $s$ would be too large to fit in BLE packets.

## 7.5 Received-Interactive Protocols

To the best of our knowledge, no proposed protocol falls into this category. As a strawman, we proposed RI-PSI. Each patient reports the beacons they received, e.g., $g^{x'}$, as their tokens. Each user uploads blinded versions of the beacons she sent by raising them to some secret $s$ (resulting in $g^{xs}$). The server blinds patient tokens by raising them to the power of a secret $t$ (resulting in $g^{x't}$). Each user periodically downloads all new patient blinded tokens, raises them to the power $s$ (resulting in $g^{x'st}$), permutes all tokens that belong to one patient, and uploads them. The server then un-blinds these tokens by raising them to the power of $1/t$ so that they become $g^{x's}$, and then match them with what the user uploaded earlier. The main downside of this protocol is the high communication/computation cost. Each user needs to download all patients tokens, raise them to some power, and then upload them.

## 7.6 Received-Server Protocols

In Received-Server protocols, the CTS conducts the matching to determine exposure events, and, thus, it needs to know the beacons each user broadcasts. The main weakness of these protocols is that there is leakage of user movement traces to the server. Several widespread protocols and already deployed smartphone applications fall into this category. I one prominent protocol, ROBERT [64], the beacons are generated by the server, and then distributed to the users. Users query the server about exposure status in a poll manner. A similar solution has been proposed in NTK [4, 7], but the server "pushes" exposure notifications to users, rather than waiting for users' queries. In BlueTrace [17], the server provides users with a secret seed, from which the users generate the beacons they broadcast, and users are notified of exposure in a push fashion. In ContraCorona [21] and TraceSecure [18], users generate random seeds from which they derive the beacons they broadcast, and all users report the secret seeds to the server. In TraceSecure, the server notifies exposed users, whereas ContraCorona presents two options: 1) the server publishes a list of all potentially infected seeds; 2) the server allows users to query for a given seed and answers whether the seed has been marked as at risk. Such variations do not affect their main properties.

## 7.7 Agreed-User

In Report-Agreed protocols, tokens are unique to each user-to-user contact event, and both users must "agree" on the token for the contact to be valid, which provides better resiliency.

In Pronto-C2 [16], users generate beacons in the form $g^a$, where $g$ is the generator of an elliptic curve group of prime order, and shared by all protocol participants, and $a$ is a random secret that changes periodically. Users upload the beacons to a shared bulletin board managed by the server and receive an address on the bulletin associated with each uploaded beacon. Users broadcast bulletin addresses rather than beacons, and contact tokens are produced in a Diffie-Hellman fashion: users retrieve the beacon associate with each received bulletin address and compute the DH token $K = H(g^{ab})$, where $b$ is the secret of the user who received the beacon. Patients report DH tokens collected, and regular users download reported DH tokens and check their contacts for exposure.

In an alternative version on Pronto-C2, named Pronto-B2 [16], users exchange plain random secrets $A$ and $B$ instead of using Diffie-Hellman style beacons. When a user is tested positive, the user uploads to the server tokens in the form $H(A \parallel B)$. Then, each user downloads the reported tokens from the server and computes the matching. This deisgn, however, leaks movement traces to Server+Psv, since an eavesdropper can also compute $H(A \parallel B)$.

## 7.8 Agreed-Interactive

In DESIRE [27], patients report Diffie-Hellman tokens related to their contacts to the server. Users periodically query the server for exposure status. Since the server performs the matching, one may argue that DESIRE belongs to the Agreed-Server category. However, in DESIRE, the server should not store tokens uploaded by users after each query, thus protecting users' privacy in the situation when an adversary compromising the server. However, a malicious server can easily store the information uploaded by the users. Thus, there is no additional privacy protection against a malicious server when comparing with a Server-Matching design,.

## 7.9 Agreed-Server

A simple change to DESIRE [27] where users periodically upload tokens and the server stores them for matching creates an agreed-server protocol. We call this S-DH. In S-DH (as in DESIRE), a user who broadcasts $g^x$ and receives $g^{x'}$ uploads $H(g^{xx'} \parallel 0)$ when $g^x < g^{x'}$, and $H(g^{xx'} \parallel 1)$ when $g^{x'} > g^x$, ensuring that the uploaded information from two non-patient users cannot be linked. When a user becomes a patient, it would upload $g^{xx'}$, enabling the server to match. Note that what a user uploads before becoming a patient can be viewed as a commitment of the value $g^{xx'}$; thus a user cannot decide to report new encounters after becoming a patient.

# 8 MORE RELATED WORK

## 8.1 Related work on survey and analysis

In addition to many the PCT protocols described in the literature, there are also several surveys classifying and comparing contact tracing protocols and applications. Some of the attacks and privacy concerns have been discussed by those surveys. Our paper, however, is unique in several aspects. First, we identify the two main design dimensions for PCT protocols, and group PCT protocols into nine categories. We show that many properties are largely determined by these two choices. Second, we systematically enumerate the possible privacy leakages, resulting in more comprehensive coverage of the privacy properties. Third, most of our analysis of resiliency attacks and defenses is missing in the literature. Fourth, due to our systematic approach, we are able to identify new protocols that offer desirable combination of properties.

Legendre, et al. [50] presented an overview analysis of the privacy and security risks of proposed contact tracing protocols. They classify the contact tracing apps into three categories: Mobile Operator Contact Tracing, Location-based Contact Tracing and Proximity-based Contact Tracing. Then they define a threat model with 3 privacy risks (Health Status Privacy, Location Privacy, Social Graph Privacy) and 3 cyber security risks (False Alarms, Passive Disruption, and Active Disruption ) and use these risks to evaluate the protocols. We list the potential risks of the apps and the design goals of this paper. For the proximity-based approach, they only analyze DP3T and ROBERT.

In a recent survey paper [6], Ahmed et al. categorize contact tracing protocols into centralized, decentralized and hybrid based on which entities is responsible for matching computation (encryption, decryption, match, and alert). Hybrid is in the middle, where the risk analysis and notifications should be the responsibility of the centralised server, and TempID generation and management remain decentralised handled by devices to ensure privacy and anonymisation. (Most likely based on whether the protocol uses Diffie Hellman, hence DESIRE, Epione, and ConTra Corona are hybrid). This survey discusses a total of 15 apps; four of them are centralized apps, 8 are decentralized app, and three hybrid apps. The analysis of each app focuses on the beacon generation, matching method, and privacy/security enhancement. This paper also discusses 8 properties: replay/reply (forwarding in our paper), wireless tracking (movement trace focus on sniffing), location confirmation (movement traces focus on psv+server), enumeration, DoS attacks, linkage, carryover, and social graph. In addition, the paper also discuss the performance properties: battery, os compatibility, content withdraw, and transparency.

Krishnan et al. [48] divide the protocols into government deployments and academia/private sector deployments in their work on contact tracing analysis . They classify the academic approaches into decentralized, centralized, GPS included and hybrid. They also identify several design challenges: security of location data, user privacy irrespective of infection status, tracing multi-hop transmission, security and trust model, and server implementation. For the attacks, they discuss ten attacks, most of which are resiliency and privacy. They also also discuss the political risk which is beyond an academic analysis.

Researchers in Fraunhofer Aisec [7] reviewed the most prominent European approaches, DP-3T, the German variant NTK of PEPP-PT, and its closely related approach, ROBERT. They do not classify these three approaches but discuss their difference with respect to several features: goals and attacker models, security properties (Sybil attack/trolling, Authorized test result reporting and Authenticity of notifications to at-risk persons), and privacy properties (De-anonymization of normal, at-risk and infected users, information gain from backend data - in centralized, and Location tracking of users). The paper also discusses scalability properties.

## 8.2 Location-based solutions

Rather than PCT, other projects are working on location-based solutions that do not require users to exchange data but rely on users' location information to detect contact events. Typically, the server collects location data (like GPS) and computes the matching to identify potential exposure. In one solution, the cellular carrier collects users' location information and estimates the exposure events, and does not require the user to download an app. However, the accuracy is around 30 meters [2], which is difficult to use for contact tracing directly. This approach leaks private information to the government or carrier, and the users cannot opt-out easily. This approach is deployed in Israel [24] and works more like a surveillance system for the citizens and makes it possible to analyze locations' density. [64, 68].

This second approach requires users' collaboration. Each user installs an application on the smartphone, which periodically records time and location (using GPS), and reports the stored movement traces to a server. The server then matches the movement traces to discover exposure events by checking whether a user was at the same time and place as a diagnosed patient. When using GPS, the accuracy is typically within a 5 meters range (16 ft) [1]. As an example of this, Iceland launched the "Rakning C-19" app, which collects users' GPS information to support more traditional contact tracing. According to an MIT report [3], by May 11th, 2020, the app has the largest penetration rate of all contact trackers in the world, with more than 38% of Iceland's population (364,000) using it. However, the effect is rather limited and needs a much higher user percentage to have a significant impact. Similar GPS solutions are proposed in [20, 62, 67], in which the app collects the GPS data of the users, and encrypt them in order to hide the location information of the users, and upload them to the sever. The server matches the different users encrypted location information and report the potential contact event exposure between patients and the normal users.

The third approach requires third party collected data. Governments or the health authorities can request third parties (e.g., groceries, restaurants, hotels) to provide specific data, such as CCTV footage, credit card usage information, or check-in/check-out dates to infer users' movements. The server then aggregates the data and computes the exposure events. An example of this approach is the method used in South Korea during the early stage of the Covid-19 outbreak to analyze people's movements, which takes just 10 minutes to track infected patients' movements [44]. A similar approach is used in China, where massive multinational technology companies such as WeChat and Ant Finance can track customers' digital payment information to analyze their movement [58].

## 8.3 Using Other Techniques for Contact Tracing

Additional solutions that rely on alternative technologies, such as WiFi, ultrasound, cellular network, rather than Bluetooth, have been discussed in previous work [6, 54, 79]. In the survey paper [32], the authors first evaluate seven popular PCT and Location-based contact tracing protocols, and then list several possible attacks and the countermeasures by the apps. Different from other survey papers, this work analyzes phone-level potential attacks, such as Resource drain attack, Screen lock attack or ransomware, and analyzes Bluetooth specific attacks, that is, Bluejacking, Bluebugging. Location-based contact tracing techniques and studies about potential digital surveillance systems are presented in several papers [43, 50, 66]. Beyond the technical aspects, some surveys evaluate implementation issues, user mental concern, and hardware impact which is rarely covered by others [6, 47, 51, 68, 71, 72, 74]. There are surveys listing additional privacy and resiliency vulnerabilities [29, 56]. Some work focuses on the analysis of specific protocols, like Google/Apple Exposure Notification (GAEN) by [41], and DP3T by [78]. Studies about acceptability of the apps among users have also been carried out [9].

## 9 CONCLUSIONS

We have introduced a framework for analyzing and comparing Proximity-based Contact Tracing (PCT) protocols. We classify all existing PCT protocols into nine categories based on two dimensions: "what do patients report" and "which party performs the matching" to discover exposure events. We then identified desirable Privacy, Utility, Resiliency, and Efficiency properties for PCT protocols, and finally analyzed these properties for protocols in the nine categories. This systematic approach enable us to understand the strengths and weaknesses of design choices made in existing PCT protocols as well as identifying promising new protocols.

For many countries, the society is unlikely to agree on contact tracing in time for such technologies to be deployed quickly and widely enough to play a significant role in fighting COVID-19.

However, we believe that it is important for us to understand the tradeoffs of different PCT protocols and reach consensus so that such technology can be quickly deployed when needed in the future.

Like the SARS coronavirus of 2003 [37] and the Middle East Respiratory Syndrome coronavirus of 2012 [13], COVID-19 is believed to have originated from coronavirus in bats [49]. That makes COVID-19 the third such respiratory-based zoonotic virus spillover event in the last 20 years. Clearly, this is not going to be the last time such an event occurs, and society should be better prepared for the inevitable future spillover events. When combined with vigilance from the public health authority and society, effective contact tracing technology may be able to contain or eradicate similar pathogens in the future, thus preventing or effectively mitigating pandemics like the disastrous COVID-19 pandemic. By providing both the design dimensions and extensive analysis of desirable properties of PCT protocols, we hope that this paper can contribute to the understanding and consensus-building on PCT protocols.